\documentclass[journal]{IEEEtran}
\usepackage{xcolor}
\usepackage{cite}
\usepackage{float}
\usepackage{amsmath,amssymb,amsfonts}
\usepackage{algorithmic}
\usepackage{booktabs} 
\usepackage{graphicx}
\usepackage{textcomp}
\def\BibTeX{{\rm B\kern-.05em{\sc i\kern-.025em b}\kern-.08em
    T\kern-.1667em\lower.7ex\hbox{E}\kern-.125emX}}
\begin{document}
\title{Functional Connectome Fingerprinting  Using Convolutional and Dictionary Learning}
\author{Yashaswini and  Sanjay Ghosh, \IEEEmembership{Senior Member, IEEE}
\thanks{This work is supported by the Faculty Start-up Research Grant (FSRG), IIT Kharagpur, awarded to Dr. Sanjay Ghosh.}
\thanks{Yashaswini  and Sanjay Ghosh are with Department of Electrical Engineering, Indian Institute of Technology Kharagpur, WB 721302, India (Email: \textcolor{blue}{sanjay.ghosh@ee.iitkgp.ac.in}).}
}

\maketitle

\begin{abstract}
Advances in data analysis and machine learning have revolutionized the study of brain signatures using fMRI, enabling non-invasive exploration of cognition and behavior through individual neural patterns. Functional connectivity (FC), which quantifies statistical relationships between brain regions, has emerged as a key metric for studying individual variability and developing biomarkers for personalized medicine in neurological and psychiatric disorders. The concept of subject fingerprinting, introduced by  Finn et al. (2015) , leverages neural connectivity variability to identify individuals based on their unique patterns. While traditional FC methods perform well on small datasets, machine learning techniques are more effective with larger datasets, isolating individual-specific features and maximizing inter-subject differences. In this study, we propose a framework combining convolutional autoencoders and sparse dictionary learning to enhance fingerprint accuracy. Autoencoders capture shared connectivity patterns while isolating subject-specific features in residual FC matrices, which are analyzed using sparse coding to identify distinctive features. Tested on the Human Connectome Project dataset, this approach achieved a 10\% improvement over baseline group-averaged FC models. Our results highlight the potential of integrating deep learning and sparse coding techniques for scalable and robust functional connectome fingerprinting, advancing personalized neuroscience applications and biomarker development.
\end{abstract}


\section{Introduction}
\label{sec:introduction}
Functional connectome fingerprinting (FCF) has emerged as a powerful approach for mapping and understanding the unique patterns of brain connectivity that characterize individuals \cite{van2013wu_et}, \cite{finn2015functional_et,tipnis2020functional_et}. The human brain operates as a dynamic network \cite{calhoun2016time} with distinct communication pathways between regions that vary between individuals \cite{allen2014tracking}. Analyzing these functional connectivity patterns enables the creation of individualized brain maps, offering critical insights into cognition, behavior, and emotion \cite{amico2018quest}. These advances have significant implications for clinical neuroscience, facilitating personalized treatment strategies for conditions such as Alzheimer’s disease \cite{zhou2024multi}, autism \cite{kunda2022improving}, and schizophrenia \cite{zhu2024temporal_et}, while enhancing our understanding of how brain networks influence mental health.
Furthermore, it helps in forensic neuroscience, aiding in applications such as knowledge verification and truth assessment. As the ability to map and analyze individual brain connectivity continues to advance, it is poised to transform personalized neuroscience and cognitive enhancement.
%
Functional connectome \cite{van2013wu_et,glasser2013minimal_et, glasser2016multi_et,damoiseaux2006consistent_et,ghosh2023joint,ghosh2023graph} provides a network-level perspective, capturing intricate interactions between brain regions during cognitive tasks \cite{cole2013multi} or emotional processing. 

%
The foundational work by Finn et al. \cite{finn2015functional_et} established that functional connectivity patterns are unique to individuals and can serve as reliable fingerprints. Using resting-state fMRI data, their study showed that intrinsic connectivity networks could accurately differentiate subjects, emphasizing the stability of functional connectomes over time.
The authors in \cite{cai2019refined_et} introduced a refined measure of functional connectomes that improved the robustness of individual identification. Their work highlighted the importance of denoising strategies and feature selection in improving fingerprinting performance.
A degree normalization step was applied to each functional connectome in \cite{chiem2022improving} to improve fingerprinting performance.
Most recently, a study by Lu et al. \cite{lu2024brain_et} leveraged high inter-subject variability features to optimize fingerprinting accuracy, suggesting that certain regions of the brain are more informative for identification than others.

Beyond conventional functional connectivity analysis, researchers have explored alternative computational methods for FCF. \cite{hannum2023high} et al.  applied high-accuracy machine learning techniques, showing that sophisticated classification models could enhance fingerprinting and cognitive state decoding. 
Cai et al. \cite{cai2021functional_et} proposed a deep learning method based on autoencoders that could improve both subject identification and prediction of cognitive functions.
In \cite{wang2024brain}, a brain recognition fingerprinting method based on graph contrastive learning was introduced. That unsupervised machine learning approach encodes brain function coupled with cognition as captured by longitudinal rs-fMRI and cognitive testing.
A large-scale study by Ogg et al. \cite{ogg2024large} demonstrated the feasibility of generalization and transfer learning in functional connectome fingerprinting, emphasizing scalability in neuroimaging applications.
The application of FCF has expanded beyond individual identification to lifespan analysis, clinical diagnostics, and behavioral prediction. The study in \cite{st2023functional_et} examined how FCF evolves across the lifespan, revealing age-related changes in connectome stability.
The work on connectome identifiability in Alzheimer’s disease in \cite{stampacchia2024fingerprints_et} suggests that alterations in functional connectivity could serve as biomarkers for neurodegeneration.
analyzed Lately, the reproducibility and precision of connectome-based fingerprinting  underscoring its potential in behavioral prediction and psychiatric research was analyzed in \cite{ramduny2025connectome}.


\begin{figure*}
    \centering
    \includegraphics[width=0.9\linewidth]{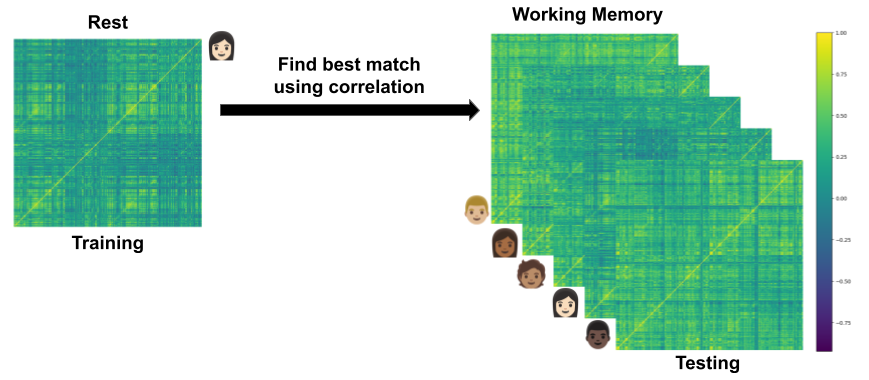}
    \caption{\textbf{Illustration of the functional connectome-based fingerprinting process}. The correlation matrix of a participant's resting-state fMRI session is compared with matrices from task-based fMRI sessions using Pearson correlation. The subject with the highest similarity score is assigned as the match. Accuracy is determined by whether the predicted identity matches the true identity.}
    \label{fig2}
\end{figure*}

In this paper, we focus on functional connectome fingerprinting and explore how it can be achieved across different cognitive tasks. The key contributions of this work are as follows:







\begin{enumerate}
    \item \textbf{Convolutional refinement of residual connectome:} We propose a novel framework that utilizes a convolutional autoencoder to learn shared connectivity structures across subjects. By subtracting the reconstructed connectome from the original, we obtain residual connectomes that highlight individual-specific features while suppressing common or task-driven variance.

    \item \textbf{Sparse dictionary learning for individual-specific decomposition:} We apply sparse dictionary learning to the residual connectomes, enabling decomposition into compact and interpretable components. These components isolate subject-specific patterns from shared group-level structures, facilitating better identifiability.

    \item \textbf{Cross-condition fingerprinting and identifiability analysis:} We perform connectome-based fingerprinting across both resting-state and task-based conditions to assess the preservation of individual-specific information. The identifiability performance is rigorously evaluated using nonparametric permutation testing to ensure statistical significance.

    \item \textbf{Functional connectome domain for efficiency and compactness:} Instead of operating on high-dimensional fMRI BOLD time series data, we perform fingerprinting in the functional connectome domain. This lower-dimensional representation captures meaningful brain connectivity patterns, allowing for more compact models, reduced computational cost, and improved scalability.
\end{enumerate}

 Rest of the paper is organized as follows. Section \ref{sec:related}  presents a detailed review of related works in functional connectome fingerprinting. In Section \ref{sec:method}, we provide the dataset and proposed method.  Section \ref{sec:result} evaluates and analyzes experimental results on subject identification across various tasks. Finally, we conclude our work in Section \ref{sec:conc}.

\section{Related Work}
\label{sec:related}
\subsection{Fingerprinting from Functional Connectome}

\subsubsection{Classical Machine Learning Methods}
Baldassarre et al. \cite{baldassarre2012individual} showed that individual differences in brain functional connectivity during resting state can predict how well individuals perform on a novel perceptual task. The experimental finding established that individual differences in performing novel perceptual tasks can be related to individual differences in spontaneous cortical activity. Airan et al. \cite{airan2016factors} studied to explore the key factors which affect the characterization and localization of inter-individual differences in functional connectivity using MRI. The robustness
of functional connectivity metrics for EEG-based personal identification was evaluated in \cite{fraschini2019robustness}. An improved fingerprinting by silencing indirect
effects while constructing functional connectome from raw EEG data was shown in \cite{jin2019extracting}. 
Brain fingerprinting using spectral analysis of functional connectomes was introduced in 
\cite{ferritto2023brain}, \cite{miri2023brain}. Brain sructure-function dependencies as informative features for  fingerprinting was demonstrated in  \cite{griffa2021structure,griffa2022brain, kotoski2024inter}.
%
Individual identification and prediction of higher cognitive functions were also performed via fingerprinting of dynamic brain connectivity patterns \cite{liu2018chronnectome} \cite{gao2021brain_et} \cite{ghaffari2025dynamic}.
FCF was also applied for predicting of cognitive decline in Parkinson’s patients \cite{stampacchia2025connectome_et} and for diagnosis of clinical migraine  \cite{esteves2024multilevel_et}. 

\subsubsection{Deep Learning Methods}
One of the first works on deep learning based brain fingerprinting was introduced in \cite{cai2021functional_et}. Cai et al. proposed a two-stage deep learning framework for task specific person identification from fMRI BOLD time series. The core idea was to enhance the individual uniqueness based on an autoencoder network. The intermediate fMRI time-series was then used to construct functional connectme. In the second stage, the functional connectmes are refined using a sparse dictionary learning to finally use for individual identification \cite{cai2021functional_et}.
The capacity of transfer learning in neuroimaging for large-scale functional connectome 
fingerprinting was explored in \cite{ogg2024large}. Inspired by speaker recognition research, Ogg et al. proposed to scale up functional connectome fingerprinting as a neural network pre-training task to learn a generalizable representation of brain function \cite{ogg2024large}.
A canonical correlation analysis (CCA) using contrastive graph learning was proposed in \cite{wang2024brain} for encoding the correlation between brain functional connectivity and cognitive measurements at individual-level. The personalized brain-cognition fingerprints characterizing brain functional differences could  reflect the unique neural and cognitive landscapes of each person. The study included healthy individuals consisting of resting-state fMRI and cognitive scores measured at multiple visits for each participant \cite{wang2024brain}. Brain graph learning for fingerprinting was further explored in \cite{balcioglu2024joint} and \cite{miri2024graph}. The principles from graph signal processing was used to learn a sparse graph structure  derived from EEG in \cite{miri2024graph}. Balcioglu et al. \cite{balcioglu2024joint} introduced a multi-task neural network architecture for joint subject-identification and task-decoding from inferred functional brain graphs constructed from raw fMRI data. 

\subsection{Fingerprinting From Brain Signal Time-series}
In parallel, researchers have also explored fingerprinting methods using various neuroimaging modalities \cite{schirrmeister2017deep_et, zhang2023overview} such as electroencephalogram (EEG), magnetoencephalography (MEG), fuctional magnetnic resonance imaging (fMRI) etc. These modalities offer high temporal resolution, providing complementary insights into the dynamic nature of brain connectivity. Studies have shown that individual-specific patterns can also be extracted from frequency-domain or source-localized representations of EEG and MEG data. 
%
Authors in \cite{schirrmeister2017deep_et} used convolutional neural networks (ConvNets) to decode movement-related information from the raw EEG for brain mapping. An EEG-based subject identification for biometric was proposed in \cite{arnau2018influence}. A data-driven analysis framework for task-independent workload assessment from EEG time-series was introduced in \cite{kakkos2021eeg_et}.
The review work in \cite{sareen2021exploring} presents the recent success of brain fingerprint from MEG. The authors in \cite{wang2019application} introduced a convolutional
recurrent neural network for individual recognition based on resting state fMRI data.
While fMRI-based fingerprinting remains dominant due to its spatial resolution, EEG and MEG-based methods hold promise for more accessible, real-time, and portable applications in clinical and cognitive neuroscience \cite{zhang2023overview}.

\section{Materials and Method}
\label{sec:method}
\subsection{Data acquisition and processing}
This study utilized the S1200 Data Release from the Human Connectome Project (HCP) \cite{van2013wu_et}, which includes behavioral and 3T MR imaging data from $1,206$ healthy young adult participants, collected between August 2012 and October 2015. A total of $889$ subjects had complete data for all four HCP 3T MRI modalities: structural images (T1w and T2w), resting-state functional MRI (fMRI), task fMRI, and high angular resolution diffusion MRI (dMRI). For this analysis, the dataset was filtered to include only participants with complete data across all task conditions: working memory, motor, language, and emotion tasks. Finally, we perform the experiments on a cohort of 339 participants.

Data acquisition was performed using a customized Siemens 3T “Connectome Skyra” scanner at Washington University in St. Louis. Advanced motion tracking systems were employed to minimize head movement. The fMRI data were collected using a whole-brain multiband gradient-echo echo-planar imaging (EPI) sequence optimized for imaging quality. Task fMRI runs for the working memory, motor, language, and emotion tasks were acquired on separate days, while resting-state fMRI runs were conducted in two separate sessions. For standardization, only fMRI data with left-to-right phase encoding were included. Detailed acquisition parameters can be found in the HCP reference manual \cite{van2013wu_et}.

Preprocessing of the fMRI data followed the HCP's minimal preprocessing pipeline (Glasser et al., 2013). This pipeline included gradient distortion correction, head motion correction, image distortion correction, spatial normalization to Montreal Neurological Institute (MNI) space, and intensity normalization. Additional steps, such as removing linear trends and band-pass filtering (0.01–0.25 Hz), were applied to reduce biophysical and other noise sources. Consistent with Finn et al. (2015) \cite{finn2015functional_et}, we do not apply spatial smoothing, as we found that this has little to no effect on identification accuracy.

To analyze the relationship between brain regions and behavior, the Glasser parcellation \cite{glasser2016multi_et} was applied, dividing the brain into 360 functional nodes. Time series for each node were extracted by averaging the time courses of all voxels within each node. These nodes were assigned to twelve functional networks: somatomotor, cingulo-opercular, orbitofrontal-affective, visual I, auditory, dorsal attention, ventral-multimodal, default, language, frontoparietal, posterior-motor, and visual II.

\subsection{Functional Connectome Generation}
For each of the \(n\) participants, with  $p$ regions-of-interest (ROIs) and $n_t$ time points, functional connectivity was calculated as a correlation matrix \(C_i \in \mathbb{R}^{p \times p}\), where \(i = 1, 2, \dots, n\). Each element \(C_i(b_1, b_2)\) represents the Pearson correlation between the time series of ROIs \(b_1\) and \(b_2\) for the \(i\)-th participant, calculated across the entire time series.  

These individual subject-specific functional connectomes were generated for all task conditions: working memory, motor, language, and emotion tasks, and resting-state. The resulting correlation matrices captured the degree of connectivity between brain regions, enabling examination of global and region-specific patterns of connectivity.  
This framework provided a detailed representation of the functional organization of human brain. This  functional organization facilitates the exploration of individual variability in connectivity patterns across different experimental conditions.

\begin{figure*}[htp!]
    \centering
    \includegraphics[width=1\linewidth]{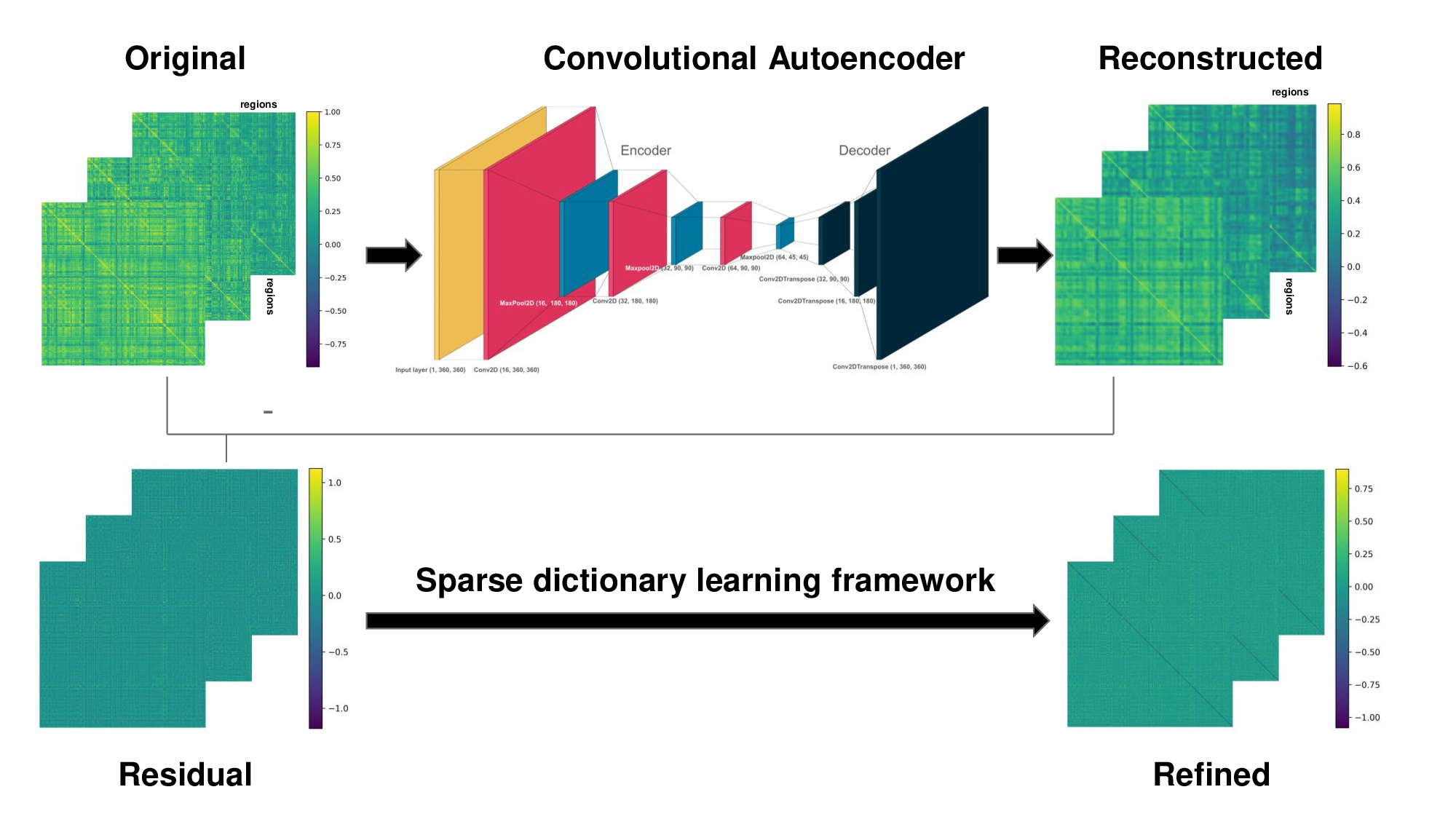}
    \caption{\textbf{ Overview of the proposed framework.} The functional connectome is first processed using a convolutional autoencoder (ConvAE) to extract a lower-dimensional latent representation and reconstruct the connectivity matrix. The residual functional connectome is then obtained by subtracting the reconstructed matrix from the original. Sparse Dictionary Learning (SDL) is applied to the residual connectome to extract subject-specific connectivity patterns while removing shared group-wise features.}
    \label{fig1}
\end{figure*}

\subsection{Proposed framework}
Our proposed framework utilizes an autoencoder designed to reduce high-dimensional data into a compressed, lower-dimensional latent vector while preserving the essential information required for accurate reconstruction of the original input. The autoencoder consists of two main parts: an encoder, \( Q(Z|X) \), which transforms the input data \( X \) into a latent vector \( Z \), and a decoder, \( P(X|Z) \), which reconstructs \( X \) from \( Z \). The key idea is that the compressed latent vector \( Z \) retains the most critical information about \( X \), enabling faithful reconstruction. 

In this study, we employed a convolutional autoencoder to capture shared features across individuals from functional connectomes. The convolutional layers in the encoder are effective at identifying local patterns and spatial relationships, which are essential for modeling the structure in functional connectivity data. By using a convolutional architecture, the autoencoder can better detect and learn common connectivity patterns, capturing both local and global aspects of brain activity that are shared among participants. Following this, we calculated the residual functional connectivity (FC) matrix, which represents the difference between the original FC matrix and the FC matrix reconstructed by the convolutional autoencoder. This residual matrix captures the unique, subject-specific information that remains after common patterns, likely influenced by task or state-related effects, are removed by the autoencoder. By isolating this residual information, we aim to focus on individual variations in functional connectivity that are not driven by shared task dynamics or general network structures. These residual FC matrices were then used as inputs to a sparse dictionary learning framework, allowing us to extract distinct, sparse components and identify representative connectivity patterns that characterize individual variability across subjects. An overview of the proposed framework is illustrated in Fig. \ref{fig1}.

Let \( C_i \) be the functional connectome for the \( i \)-th subject, where \( C_i(b_1, b_2) \) represents the Pearson correlation between regions of interest (ROIs) \( b_1 \) and \( b_2 \) for the \( i \)-th subject across the entire time series. To refine the functional connectome, we pass \( C_i \) through a convolutional autoencoder. The autoencoder learns a compressed, lower-dimensional latent representation of the connectivity data, which is then used to reconstruct the original functional connectome. The reconstructed functional connectome is denoted as \( \hat{C}_i^{\text{reconstructed}} \). The residual functional connectome is then obtained by subtracting the reconstructed connectome from the original:
\[
\hat{C}_i = C_i - \hat{C}_i^{\text{reconstructed}},
\]

\noindent where \( \hat{C}_i \) represents the refined residual functional connectome that captures the unique, subject-specific connectivity patterns after removing common, shared patterns likely influenced by general network structure or task-related effects.
\\

\noindent 
After refining the functional connectome, we proceed by modeling the sparse representation of the residual connectomes across all subjects using Sparse Dictionary Learning (SDL). This is formulated as follows:

\[
\min_{D, X} \| Y - D X \|_F^2 \quad \text{subject to} \quad \| x_i \|_0 \leq L, \quad i = 1, 2, \dots, n,
\]
\noindent where \( Y = [e_1, e_2, \dots, e_n] \) is the matrix of concatenated edge weight vectors, \( D \in \mathbb{R}^{m \times K} \) denotes the dictionary, and \( K \) is the number of dictionary atoms. \( X \in \mathbb{R}^{K \times n} \) is the representation matrix, with \( \| \cdot \|_0 \) controlling sparsity via the \( \ell_0 \)-norm and \( \| \cdot \|_F \) representing the Frobenius norm. The dictionary learning process is implemented using the \( \text{ksvdbox13} \) method.
Finally, to obtain the refined functional connectome, we subtract the contribution of the learned dictionary atoms from the residual connectome:

\[
\hat{C}_i = C_i - \text{mat}(D x_i),
\]

\noindent where \( \text{mat}(D x_i) \in \mathbb{R}^{p \times p} \) is the correlation matrix reconstructed from the sparse representation \( D x_i \). This process allows us to extract individual-specific connectivity patterns while removing shared group-wise features. Note that the SDL model is applied to each subject's ROI network for different fMRI modalities separately.

\subsection{Individual identifiability analysis}
To explore the use of functional connectomes as fingerprints, we applied a methodology for individual identification based on the approach proposed by Finn et al. (2015) \cite{finn2015functional_et}. The goal was to determine whether the connectivity patterns from different fMRI sessions could reliably identify the same individual. Identification was performed by training the model on resting-state fMRI data and testing it on data from other task-based fMRI sessions. Specifically, we focused on three task combinations: resting-state vs. motor (rest-Mt), resting-state vs. working memory (rest-Wm), and resting-state vs. emotion (rest-Em).

For each participant $i$, we compared the correlation matrix of their target session (e.g., resting-state 1, R1) with the matrices from all participants $j$ in session 2 (e.g., motor, Mt), using the Pearson correlation coefficient to estimate similarity. The participant was assigned the label of the subject in session 2 with the highest similarity score, formally defined as:
\begin{equation}
\hat{j} = \arg\max_{j} ; \mathrm{corr}\left(\mathbf{C}_i^{(1)}, \mathbf{C}_j^{(2)}\right)
\end{equation}
where $\mathbf{C}_i^{(1)}$ and $\mathbf{C}_j^{(2)}$ denote the connectivity matrices of subject $i$ in session 1 and subject $j$ in session 2, respectively, and $\mathrm{corr}(\cdot, \cdot)$ is the Pearson correlation coefficient. If $\hat{j} = i$, the identification was considered correct, yielding an accuracy of 100\%; otherwise, the accuracy for that comparison was considered 0\%. This process was performed for all possible session combinations. An overview of the fingerprinting process is illustrated in Fig.~\ref{fig2}.

After obtaining the identification accuracy for all the participants, we performed 1000 nonparametric permutation tests to assess whether the observed accuracies were significantly above chance. For each permutation, we randomized the identities of the subjects, repeated the identification process, and recorded the accuracies. The distribution of accuracies from these permutations provided a baseline to compare against the actual identification accuracies. A significance level of $p = 0.05$ was used as the threshold for determining whether the identification accuracy was significantly greater than chance.

This approach allowed us to evaluate the reliability and robustness of functional connectomes as individual fingerprints and provided a framework for further exploration of connectome-based biomarker applications in personalized neuroscience.

\subsection{Baseline Method}
To compare our results and assess the improvements achieved by our proposed framework,
we first established a baseline model by calculating a group-averaged functional connectome
across all subjects. This group-averaged connectome was derived by computing the Pearson
correlation between the time courses of brain regions for all subjects, resulting in an average
functional connectivity matrix that represents the overall connectivity patterns across the
entire sample. To isolate subject-specific variations in connectivity, we then calculated
the residual functional connectomes by subtracting the group-averaged connectome from
each individual’s functional connectome. This step allowed us to focus on the unique
connectivity patterns of each subject, removing the common shared connectivity structure.

We then applied sparse dictionary learning to the residual functional connectomes using
the k-SVD (sparse singular value decomposition) approach. Sparse dictionary learning is
a technique that decomposes the residual connectomes into a sparse linear combination of
learned dictionary atoms, with the goal of identifying a compact representation of the data
that captures the essential connectivity features. We experimented with different values
of the sparsity parameters k (the number of atoms in the dictionary) and l (the sparsity
parameter) to explore how varying levels of sparsity affected the learned dictionary and its
ability to represent the connectivity patterns. By adjusting these parameters, we assessed
the impact of different regularization levels on the learned features and compared the
results with our proposed framework to evaluate the improvements achieved. This baseline
method served as a point of comparison for our more advanced approach, allowing us to
quantify the benefits of incorporating additional computational techniques into our model.

\section{Results}
\label{sec:result}
\subsection{Baseline Model vs. Our Model: Identification Accuracies Across Different $K$ and $L$ Values}

\begin{figure*}[htb!]
    \centering
    \includegraphics[width=0.8\linewidth]{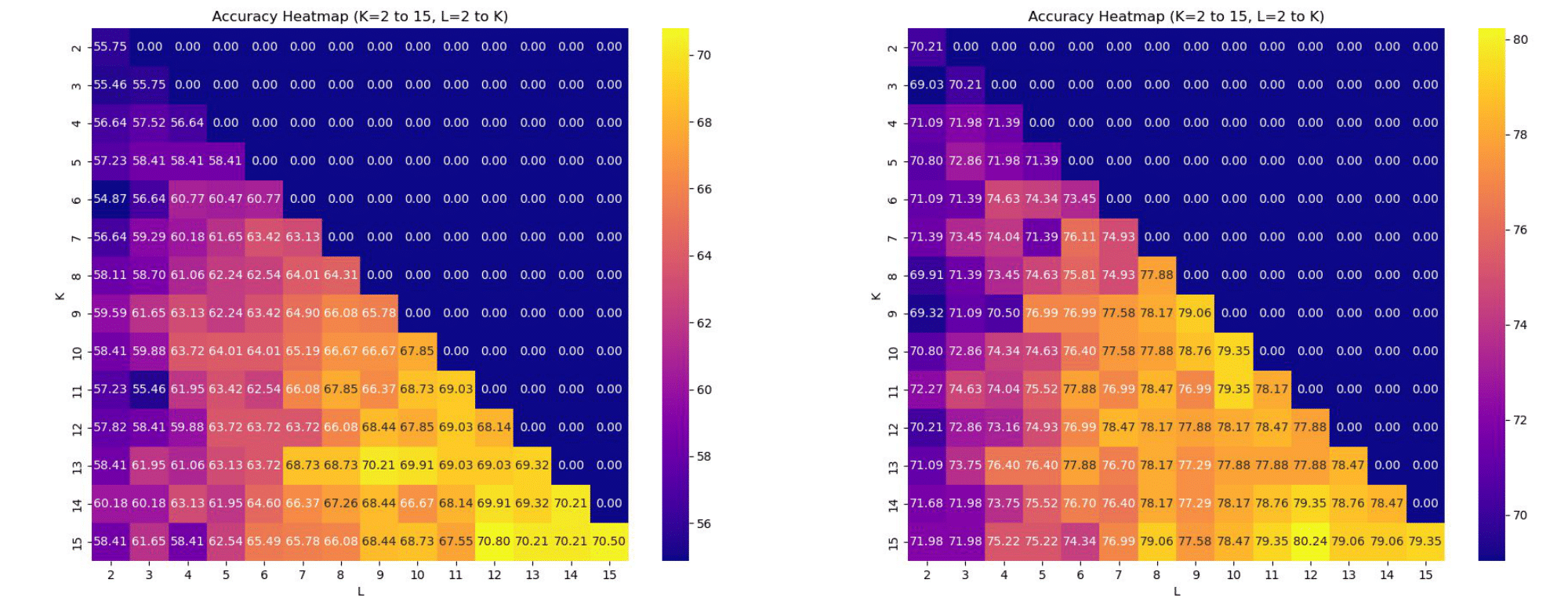}
    \caption{Motor task: Identification accuracies for different $K$ and $L$ values. Baseline model (Left) vs. Our Model (Right).}
    \label{fig:motor-kl}
\end{figure*}

\begin{figure*}[htb!]
    \centering
    \includegraphics[width=0.8\linewidth]{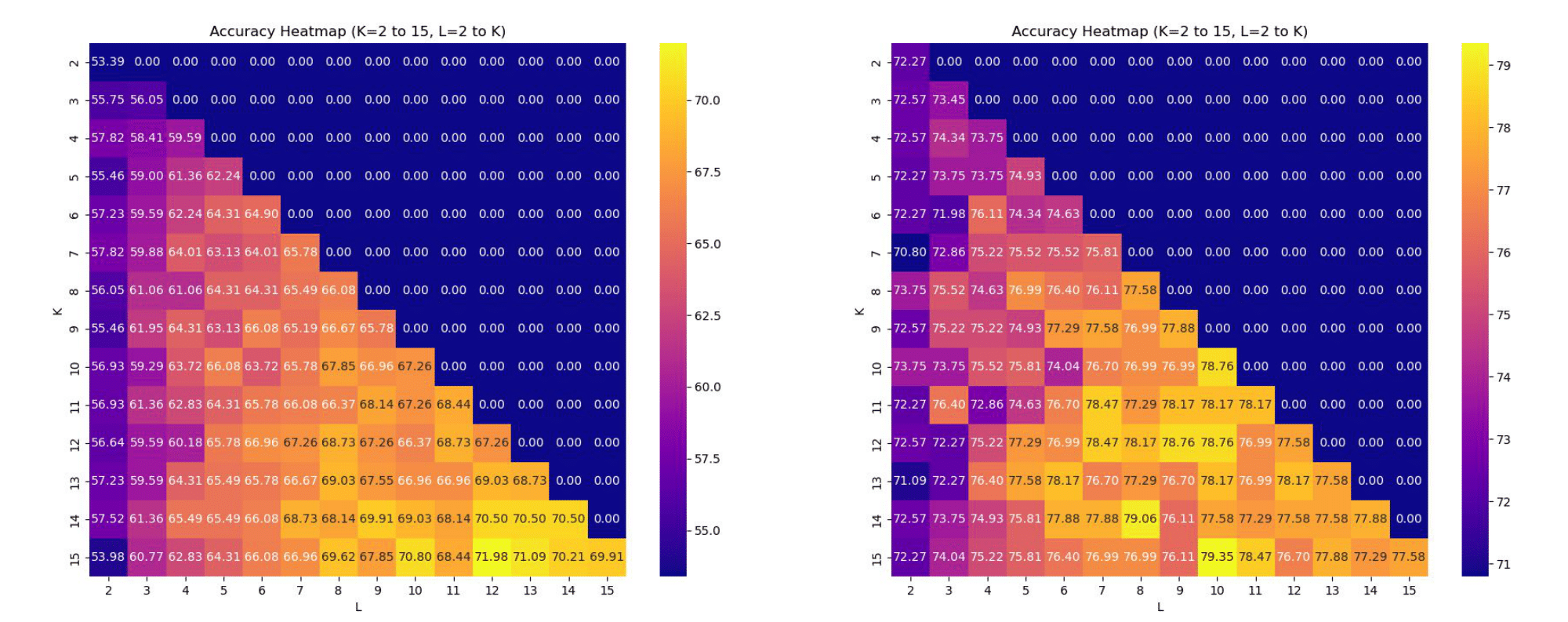}
    \caption{Working memory task: Identification accuracies for different $K$ and $L$ values. Baseline model (Left) vs. Our Model (Right).}
    \label{fig:wm-kl}
\end{figure*}

\begin{figure*}[htb!]
    \centering
    \includegraphics[width=0.8\linewidth]{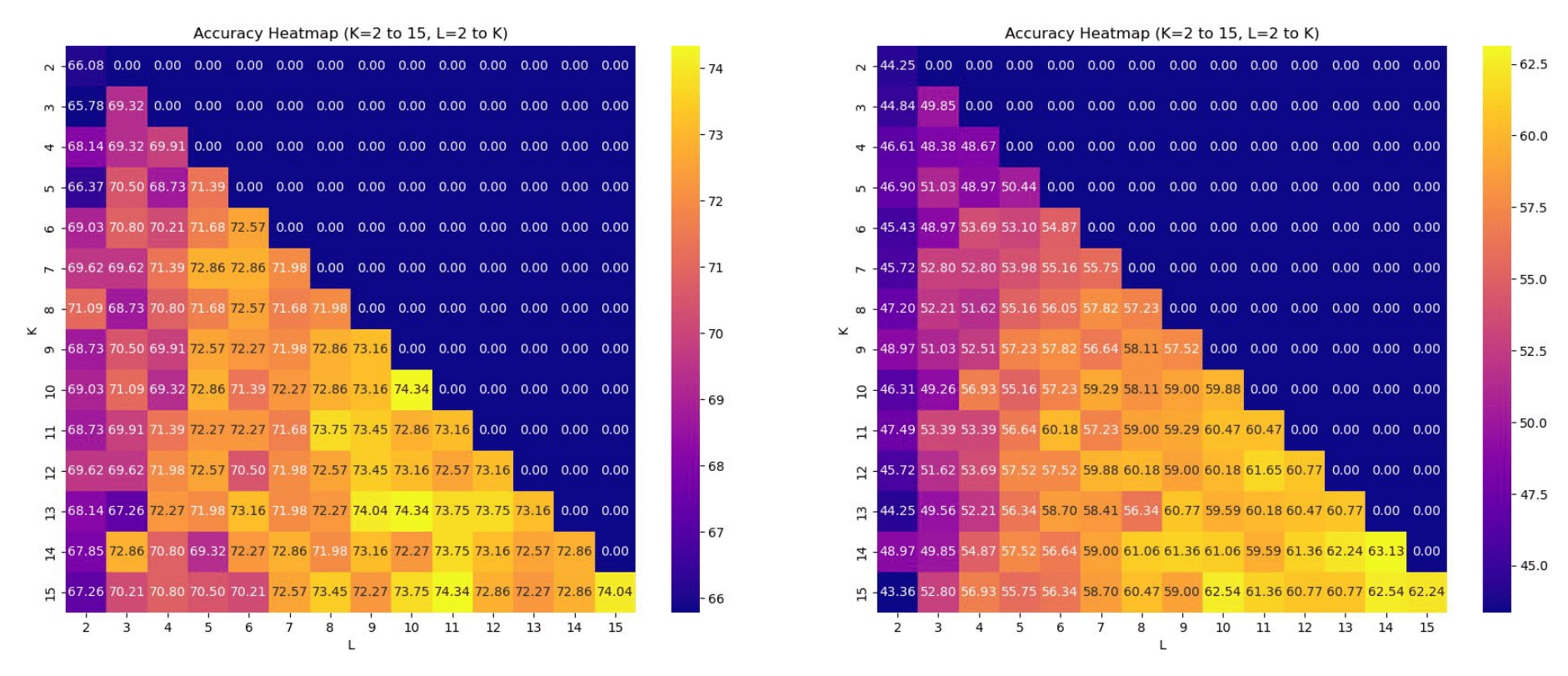}
    \caption{Emotion task: Identification accuracies for different $K$ and $L$ values. Baseline model (Left) vs. Our Model (Right).}
    \label{fig:emotion-kl}
\end{figure*}

We evaluated the identification accuracies of our model against the baseline model for motor, working memory, and emotion task-based functional connectomes. Our model was trained on resting-state data, and identification accuracy was measured for different values of $K$ (number of dictionary atoms) and $L$ (sparsity level) in Sparse Dictionary Learning. Specifically, we varied $K$ and $L$ from 2 to 15 to analyze their impact on identification performance. The results for the motor, working memory, and emotion tasks are presented in Fig.~\ref{fig:motor-kl}, Fig.~\ref{fig:wm-kl}, and Fig.~\ref{fig:emotion-kl}, respectively, illustrating the performance improvements achieved by our proposed model.  
Table 1 summarizes the results for the best-performing $K$ and $L$ values, that yielded the highest identification accuracies across the different task-based functional connectomes.

We observe an approximate 10\% increase in identification accuracy across all task modalities (motor, working memory, and emotion) for our proposed model compared to the baseline model. Specifically, we identify the optimal values of $K$ and $L$ that yield the best performance. For the motor task, the proposed model achieves a maximum accuracy of 80.24\% at $K=15$, $L=12$, whereas the baseline model attains 70.5\% at $K=15$, $L=15$. Similarly, for the working memory task, the proposed model reaches 79.35\% accuracy at $K=15$, $L=10$, while the baseline model records 71.98\% at $K=15$, $L=12$. For the emotion task, the proposed model attains 74.34\% accuracy at $K=15$, $L=11$, in contrast to the baseline model's 63.13\% at $K=14$, $L=14$.

\begin{table}[h]
    \centering
    \caption{Comparison of Finn accuracy, baseline accuracy, and our accuracy for different tasks when trained on resting state data}
    \label{tab:accuracy_comparison}
    \begin{tabular}{lccc}
        \toprule
        \textbf{Task} & \textbf{Finn Accuracy} & \textbf{Baseline Accuracy} & \textbf{Our Accuracy} \\
        \midrule
        Motor & 28.02\% & 70.80\% & \textbf{80.24\%} \\
        Working Memory & 38.35\% & 71.98\% & \textbf{79.35\%} \\
        Emotion & 19.47\% & 62.54\% & \textbf{74.34\%} \\
        \bottomrule
    \end{tabular}
\end{table}

\subsection{Subject Correlation Matrices: Before ConvAE, After ConvAE, After SDL}
\begin{figure*}[!htb]
    \centering
    \includegraphics[width=1\linewidth]{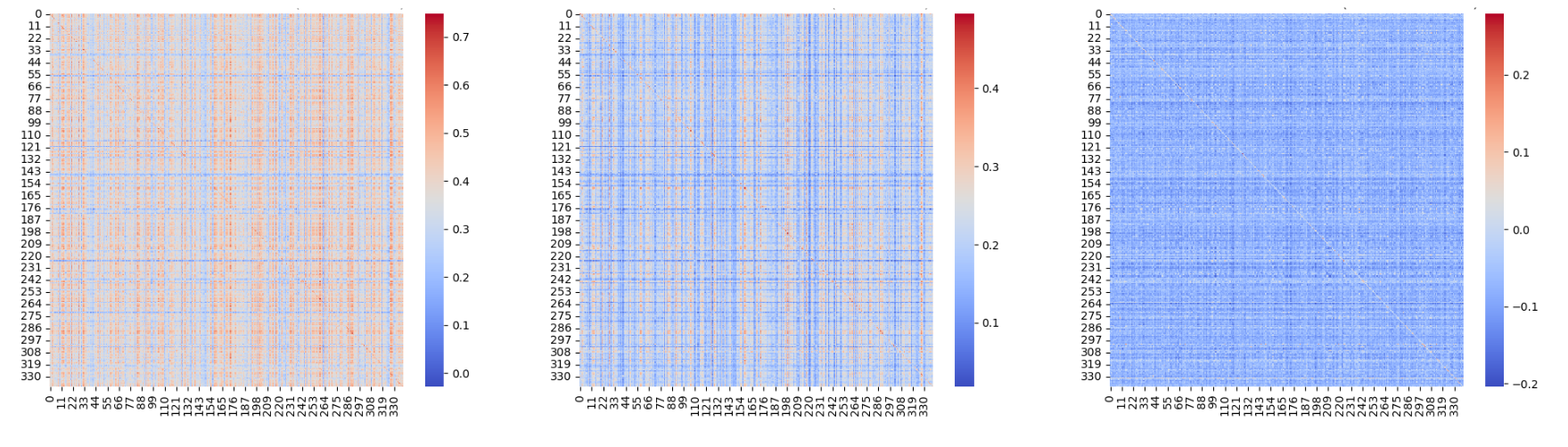}
    \makebox[0.32\linewidth][c]{(a)}%
    \hfill
    \makebox[0.32\linewidth][c]{(b)}%
    \hfill
    \makebox[0.32\linewidth][c]{(c)}

    \caption{Subject correlation matrices comparing resting-state and motor task functional connectomes across subjects.
(a) The raw correlation matrix demonstrates high similarity both along and off the diagonal, indicating limited ability to distinguish individual subjects.
(b) After applying the convolutional autoencoder (ConvAE), overall correlation values decrease, but the contrast between diagonal (matched subjects) and off-diagonal (unmatched subjects) elements increases, reflecting improved subject specificity.
(c) The subsequent application of sparse dictionary learning (SDL) further accentuates this diagonal-off-diagonal separation, highlighting more distinct subject-specific connectivity patterns and resulting in enhanced individual identification accuracy.
}
    \label{fig:rest-motor}
\end{figure*}

\begin{figure*}[!htb]
    \centering
    \includegraphics[width=1\linewidth]{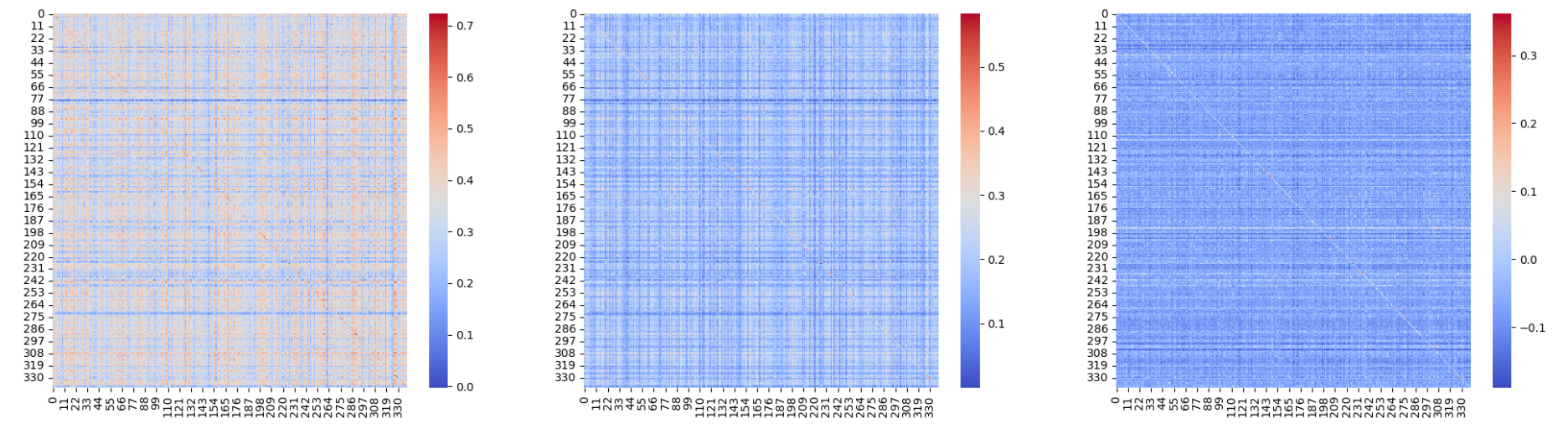}
    \makebox[0.32\linewidth][c]{(a)}%
    \hfill
    \makebox[0.32\linewidth][c]{(b)}%
    \hfill
    \makebox[0.32\linewidth][c]{(c)}
    \caption{Subject correlation matrices comparing resting-state and working memory task functional connectomes across subjects.
    (a) In the unprocessed data, high correlations are observed throughout the matrix, suggesting that common neural activity dominates and individual differences are not well captured.
(b) ConvAE processing reduces overall correlation levels and increases the contrast between matched and unmatched subjects, indicating better preservation of unique subject features.
(c) With the addition of SDL, the diagonal elements become even more prominent, and off-diagonal values are further suppressed, demonstrating the effectiveness of these combined methods in isolating and enhancing subject-specific connectivity signatures.}
    \label{fig:rest-wm}
\end{figure*}

\begin{figure*}[!htb]
    \centering
    \includegraphics[width=1\linewidth]{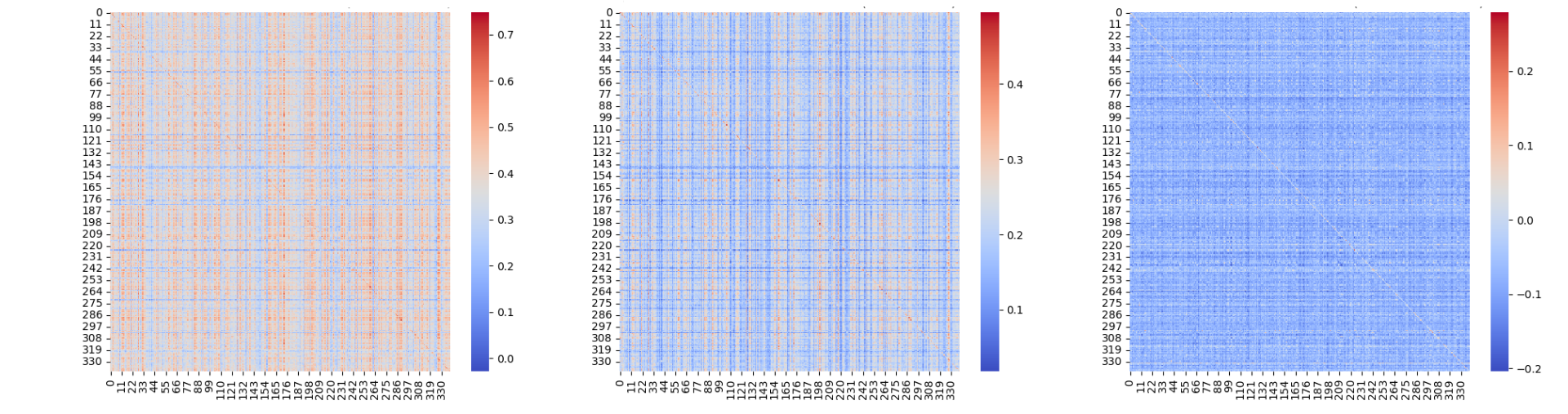}
    \makebox[0.32\linewidth][c]{(a)}%
    \hfill
    \makebox[0.32\linewidth][c]{(b)}%
    \hfill
    \makebox[0.32\linewidth][c]{(c)}
    \caption{Subject correlation matrices comparing resting-state and emotion task functional connectomes across subjects. (a) The raw matrix shows generally high and uniform correlations, making it difficult to differentiate individuals based on their connectomes.
(b) Following ConvAE transformation, the matrix displays reduced overall correlation but increased contrast, with diagonal elements more clearly standing out from the off-diagonal ones, signifying improved individual discrimination.
(c) The integration of SDL with ConvAE further sharpens this distinction, yielding a matrix where subject-specific patterns are strongly emphasized, thereby maximizing identification accuracy and demonstrating the robustness of the combined approach.}
    \label{fig:rest-emotion}
\end{figure*}

Subject correlation matrices i.e. Pearson correlation coefficient between functional connectivity across subjects between resting-state and task-based functional connectomes (motor, working memory, and emotion), highlights how the step-by-step application of ConvAE and SDL methods progressively improves individual identification accuracy. Note that row and column subject order is symmetric. Thus, diagonal elements are correlation scores from the matched subjects, while off diagonal coefficients are from the unmatched subjects. The results for all three possible task pairs—rest-motor, rest-working memory, and rest-emotion—are displayed in Fig.~\ref{fig:rest-motor}, Fig.~\ref{fig:rest-wm}, and Fig.~\ref{fig:rest-emotion}, respectively. 

In comparison to the raw cross-subject correlation coefficients, the scores generated after applying the convolutional autoencoder exhibit a significant reduction for both diagonal and off-diagonal elements. However, the convolutional autoencoder improves the contrast between diagonal and off-diagonal elements in the correlation matrix, indicating that reducing the contribution of common neural activity enhances individual discrimination. Furthermore, the combined application of the convolutional autoencoder and sparse dictionary learning (SDL) further strengthens this effect, as evident in the figures, demonstrating its effectiveness in refining subject-specific connectivity patterns and improving identification accuracy.

\subsection{Ablation Analysis}  
\begin{figure}[!htb]
    \centering
    \includegraphics[width=1\linewidth]{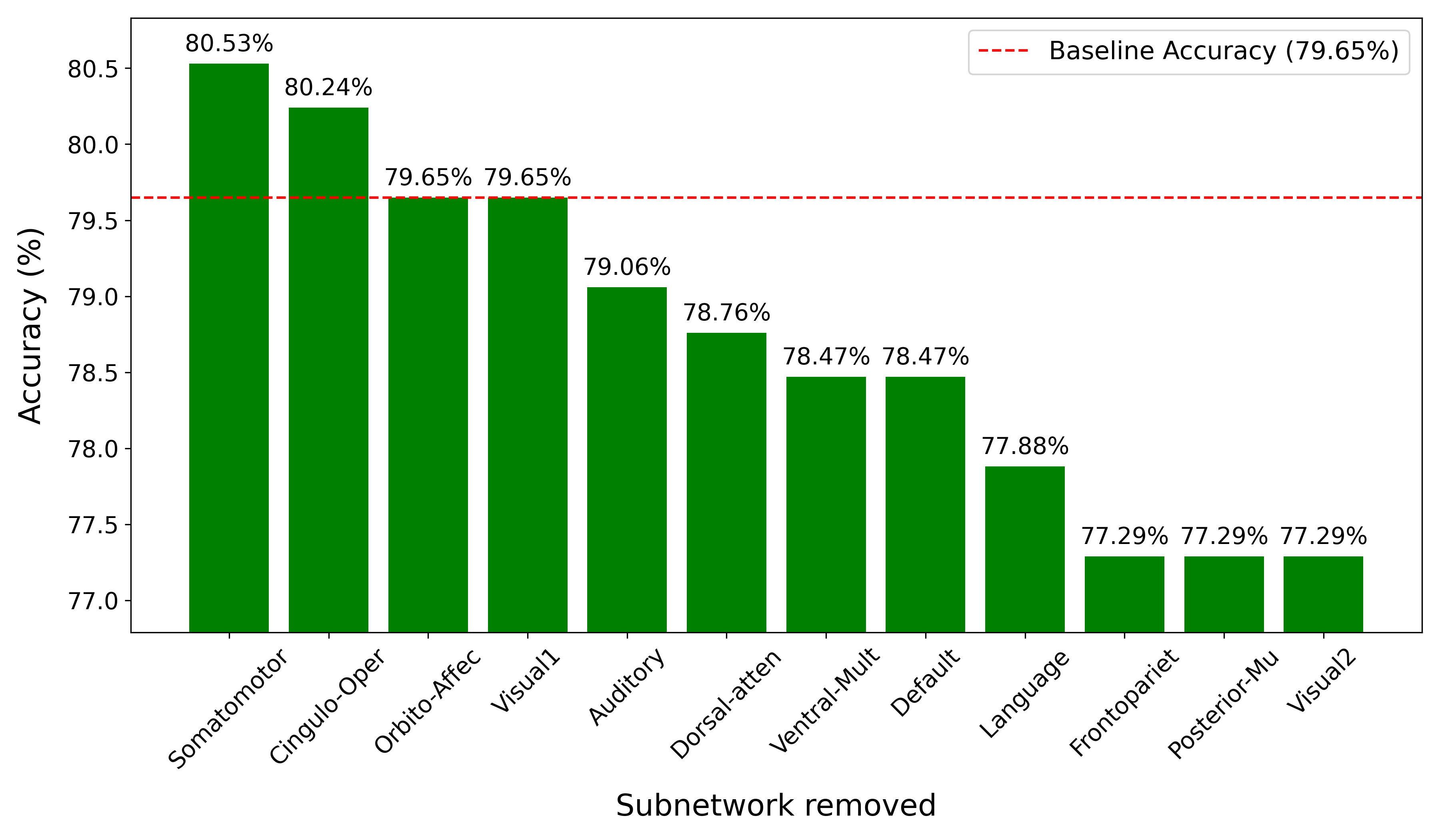}
    \caption{Motor Task}
    \label{fig:motor-ablation}
\end{figure}

\begin{figure}[!htb]
    \centering
    \includegraphics[width=1\linewidth]{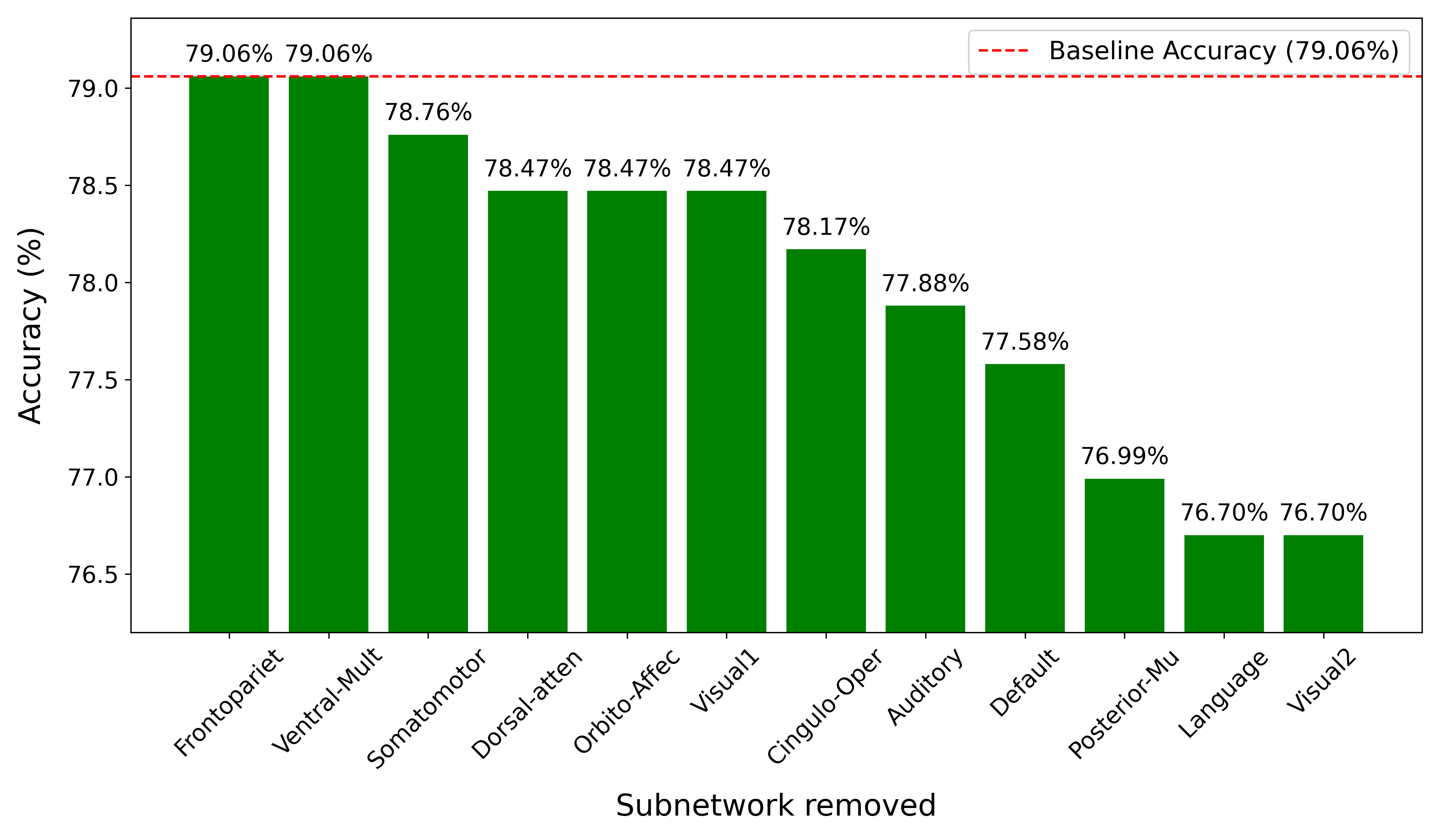}
    \caption{Working Memory Task}
    \label{fig:wm-ablation}
\end{figure}

\begin{figure}[!htb]
    \centering
    \includegraphics[width=1\linewidth]{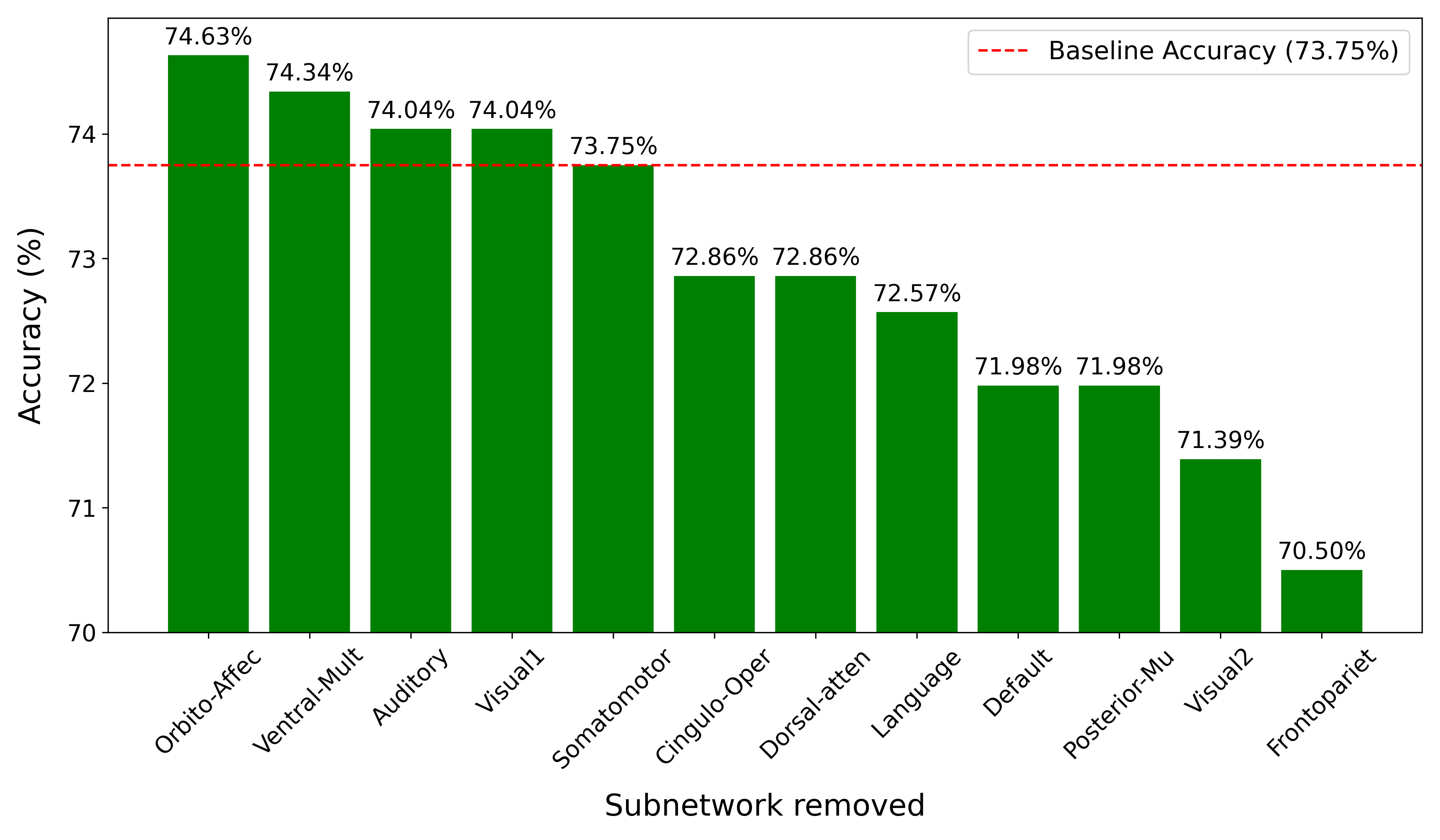}
    \caption{Emotion Task}
    \label{fig:emotion-ablation}
\end{figure}

To assess the contribution of different functional subnetworks to individual identification, we performed an ablation analysis by selectively excluding specific subnetworks and evaluating the resulting impact on identification accuracy across the motor, working memory, and emotion tasks. The results for each task are presented in Fig.~\ref{fig:motor-ablation}, Fig.~\ref{fig:wm-ablation}, and Fig.~\ref{fig:emotion-ablation}, respectively.  

For the \textbf{motor task} (Fig.~\ref{fig:motor-ablation}), the largest decrease in accuracy was observed when excluding the frontoparietal, posterior-mu, and visual2 networks. The frontoparietal network is crucial for motor planning and control, while the posterior-mu network plays a key role in sensorimotor integration. Additionally, the visual2 network likely supports visuomotor coordination, further emphasizing its relevance in motor-related tasks. In contrast, excluding the cingulo-opercular and somatomotor networks resulted in a slight increase in accuracy, suggesting that these subnetworks may introduce redundancy or noise or may not be essential for individual characterization in motor-based functional connectomes.  

For the \textbf{working memory task} (Fig.~\ref{fig:wm-ablation}), the largest accuracy decrease was observed when excluding the language and visual2 networks. The language network is critical for working memory, particularly for tasks requiring symbolic reasoning and verbal processing. Similarly, the visual2 network plays a significant role in maintaining spatial and visual details in working memory. On the other hand, excluding the auditory, cingulo-opercular, frontoparietal, or ventral-multimodal networks had minimal impact on accuracy, indicating that these subnetworks may not be primary contributors to individual differentiation in working memory-based functional connectomes.  

For the \textbf{emotion task} (Fig.~\ref{fig:emotion-ablation}), accuracy decreased most significantly when the frontoparietal and visual2 networks were excluded. The frontoparietal network is essential for emotional regulation and integration, while the visual2 network plays a role in visuospatial processing of emotional stimuli, particularly facial expressions. Interestingly, excluding the orbito-affective, ventral-multimodal, auditory, and visual1 networks resulted in slight or negligible increases in accuracy, suggesting that these subnetworks do not play a central role in individual fingerprinting for the emotion task.  

Overall, these results highlight the differential contributions of functional subnetworks to identification performance across tasks, reinforcing the importance of network-specific feature selection in functional connectome fingerprinting.

\subsection{Training on other tasks}
\begin{table}[h]
    \centering
    \label{tab:accuracy_comparison_motor}
    \caption{Comparison of baseline accuracy and our accuracy for rest state and other tasks when trained on motor task data}
    \begin{tabular}{lcc}
        \toprule
        \textbf{Task} & \textbf{Baseline Accuracy} & \textbf{Our Accuracy} \\
        \midrule
        Rest & 64.89\% & \textbf{78.76\%} \\
        Working Memory & 87.02\% & \textbf{92.03\%} \\
        Emotion & 55.45\% & \textbf{69.91\%} \\
        \bottomrule
    \end{tabular}
\end{table}

\begin{table}[h]
    \centering
    \label{tab:accuracy_comparison_wm}
    \caption{Comparison of baseline accuracy and our accuracy for resting state and other tasks when trained on working memory task data}
    \begin{tabular}{lcc}
        \toprule
        \textbf{Task} & \textbf{Baseline Accuracy} & \textbf{Our Accuracy} \\
        \midrule
        Rest & 61.94\% & \textbf{72.56\%} \\
        Motor & 83.18\% & \textbf{87.90\%} \\
        Emotion & 55.75\% & \textbf{69.32\%} \\
        \bottomrule
    \end{tabular}
\end{table}

\begin{table}[htbp]
    \centering
    \caption{Comparison of baseline accuracy and our accuracy for resting state and other tasks when trained on emotion task data}
    \label{tab:accuracy_comparison_emotion}
    \begin{tabular}{lcc}
        \toprule
        \textbf{Task} & \textbf{Baseline Accuracy} & \textbf{Our Accuracy} \\
        \midrule
        Rest & 60.14\% & \textbf{70.34\%} \\
        Motor & 55.45\% & \textbf{69.91\%} \\
        Working Memory & 55.75\% & \textbf{69.32\%} \\
        \bottomrule
    \end{tabular}
\end{table}

\section{Conclusion}
\label{sec:conc}
In this study, we introduced a novel framework that integrates convolutional autoencoders with sparse dictionary learning to enhance functional connectome fingerprinting. Our approach aims to refine individual brain connectivity representations by isolating subject-specific features while filtering out common network structures. By leveraging deep learning for feature extraction and sparse coding for efficient representation, our method provides a significant improvement in identification accuracy compared to traditional baseline models. 

Our results demonstrate an approximate 10\% increase in identification accuracy across multiple task-based functional connectomes, highlighting the robustness of our approach. Specifically, for the motor task, our model achieved a maximum accuracy of 80.24\% at $K=15$, $L=12$, while the baseline model attained only 70.5\% at $K=15$, $L=15$. Similarly, for the working memory task, our model reached 79.35\% accuracy at $K=15$, $L=10$, outperforming the baseline model, which recorded 71.98\% at $K=15$, $L=12$. In the emotion task, our framework achieved 74.34\% accuracy at $K=15$, $L=11$, surpassing the baseline accuracy of 63.13\% at $K=14$, $L=14$. These findings indicate that combining deep learning and sparsity-based techniques effectively captures individual differences in functional connectivity, leading to enhanced subject identification.

Beyond improvements in identification accuracy, the implications of our findings extend to various domains in neuroscience and medical imaging. Functional connectome fingerprinting has the potential to play a critical role in precision medicine, where understanding individual brain connectivity patterns can lead to personalized treatment strategies for neurological and psychiatric disorders. Additionally, this methodology can contribute to cognitive modeling by uncovering how functional networks differ across individuals in response to various cognitive tasks. The ability to accurately differentiate individuals based on their functional connectivity also holds promise for neurotechnological applications, such as brain-computer interfaces (BCIs), cognitive state decoding, and biomarker discovery for neurological diseases.

Despite these advancements, there are several areas for future exploration. First, while our study focused on resting-state and task-based functional connectomes, applying this framework to multi-modal neuroimaging data, including structural and diffusion MRI, could further enhance its robustness and applicability. Second, evaluating the generalizability of our method across diverse populations and clinical cohorts would be an important step toward translating these findings into real-world medical applications. Finally, exploring adaptive and hierarchical sparse dictionary learning techniques could provide a more flexible representation of individual variability in brain connectivity.

In summary, our proposed framework advances the state of functional connectome fingerprinting by effectively leveraging deep learning and sparse representation techniques to improve individual identifiability. These results underscore the potential of combining neuroscience-inspired machine learning approaches with functional connectivity analysis to deepen our understanding of individual brain organization and facilitate personalized neuroscience applications.

\bibliographystyle{IEEEtran}
\bibliography{ref}

\end{document}